\documentclass[11pt,twoside]{article}
\usepackage{asp2010}
\usepackage{graphicx}
\usepackage{amssymb}

\resetcounters

\begin{document}

\title{POPCORN: A comparison of binary population synthesis codes}
\author{J.S.W. Claeys,$^{1,2}$ S.Toonen,$^1$ and N. Mennekens$^3$
\affil{$^1$Department of Astrophysics, IMAPP, Radboud University, PO Box 9010, 6500 GL Nijmegen, The Netherlands}
\affil{$^2$Sterrekundig Instituut Utrecht, The Netherlands}
\affil{$^3$Astrophysical Institute, Vrije Universiteit Brussel, Pleinlaan 2, 1050 Brussels, Belgium}}

\begin{abstract}
We compare the results of three binary population synthesis codes to understand the differences in their results. As a first result we find that when equalizing the assumptions the results are similar. The main differences arise from deviating physical input.
\end{abstract}

\paragraph{Introduction}
Binary population synthesis (BPS) codes enable simulations of the entire evolution of a large number of binary systems. BPS codes are aimed for the study of the formation and evolution of astronomical populations, e.g. novae, X-ray binaries and type Ia supernovae (SNeIa). However, considerable differences arise in the predictions of e.g. the SNeIa rate. In this research we compare the results of three different BPS codes and investigate the importance of the different assumptions in these codes.

\paragraph{Method}
To compare the results  of  various BPS codes, we look at two populations at their time of formation: \emph{Single WD systems with a non-degenerate companion} and \emph{double WD systems}. We analyze the evolution paths towards the population of binary WDs, in order to have a complete -within the limits of this project- overview of the similarities and causes of the differences in the results of the BPS codes. As a first test we equalize the assumptions. These assumptions are not necessarily considered to be realistic. We simulate a grid with the same initial distributions (in primary mass, mass ratio, separation). We assume conservative mass transfer to all types of stars and for the common envelope evolution we use the prescription based on \cite{Webbink84}. We do not consider magnetic braking, tides and wind accretion. For this paper we only consider the formation of WDs with a mass higher than 0.48 M$_{\odot}$. Three different BPS codes are compared:
\begin{itemize}
\item \emph{Binary$\_$c}: Based on \citet{Hurley00,Hurley02}, with updates described in \cite{Izzard06} and Claeys et al. (in prep.)
\item \emph{SeBa}: Based on \cite{Portegies96}, with updates described in \cite{Nelemans01} and \cite{Toonen12}
\item \emph{Brussels code}: Based on \cite{DonderVanbeveren04}, with updates described in \cite{Mennekens10}
\end{itemize}

\paragraph{First results} 
\begin{figure}
\includegraphics[height=.15\textheight]{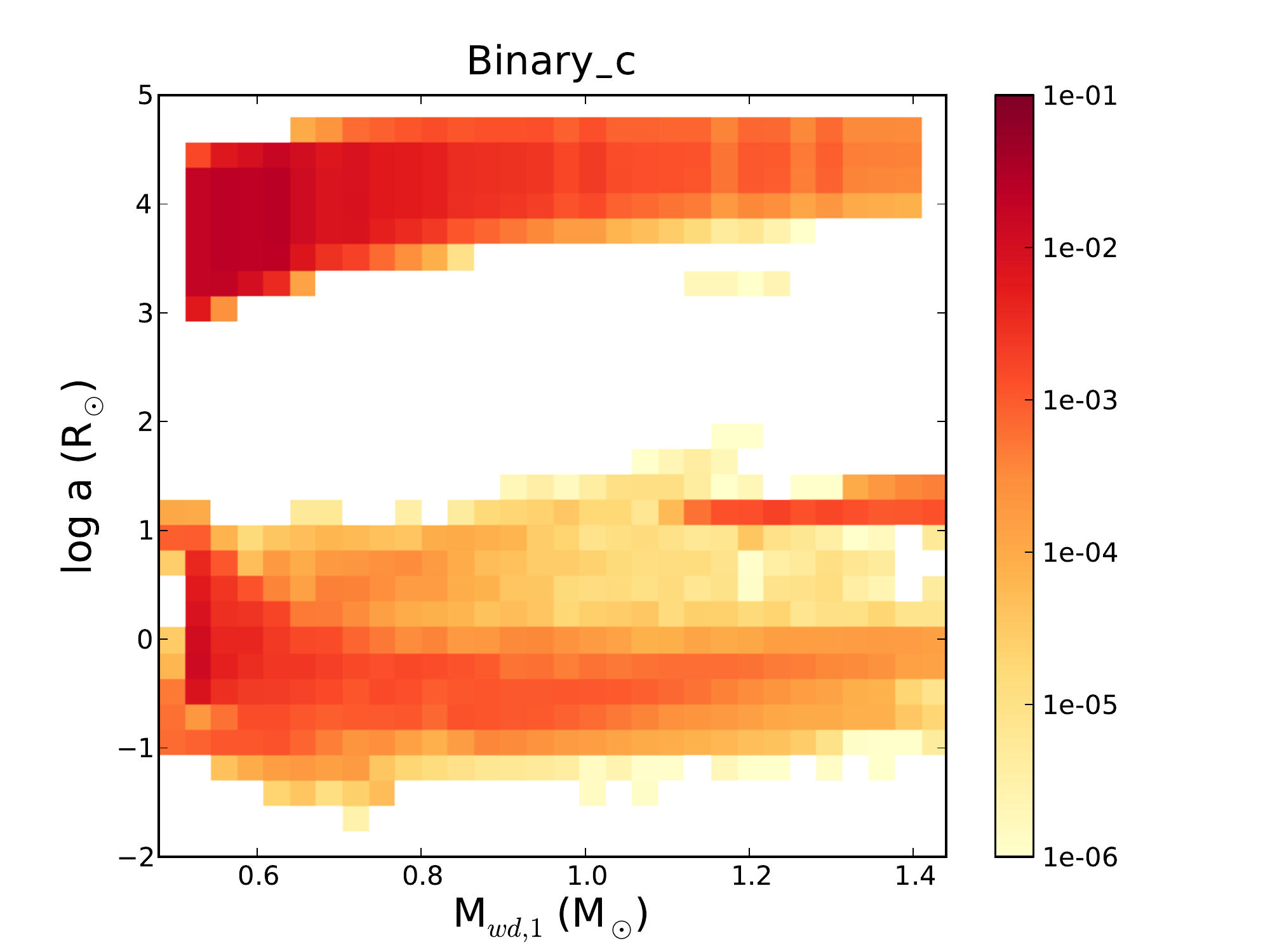}
\includegraphics[height=.15\textheight]{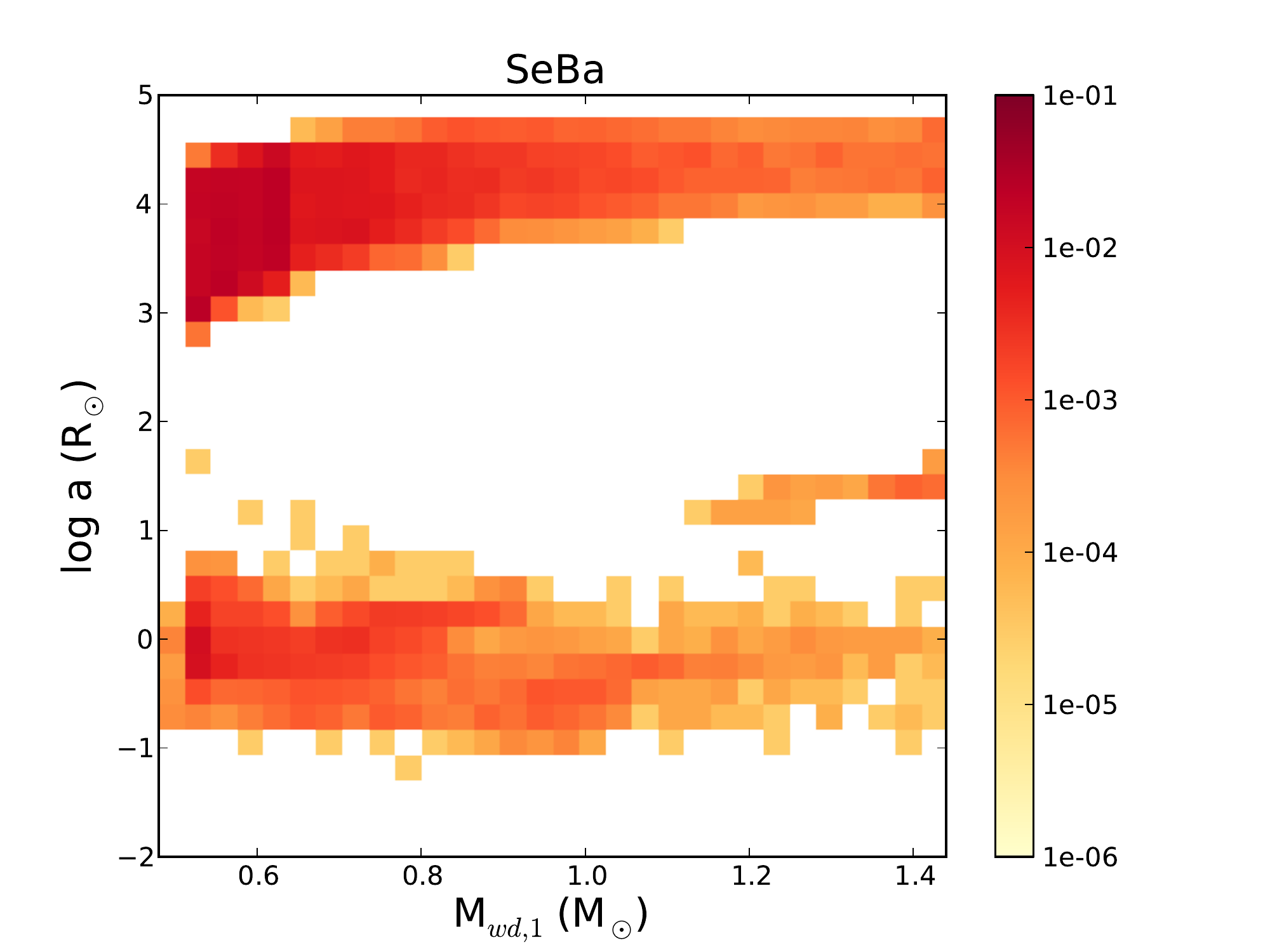}
\includegraphics[height=.15\textheight]{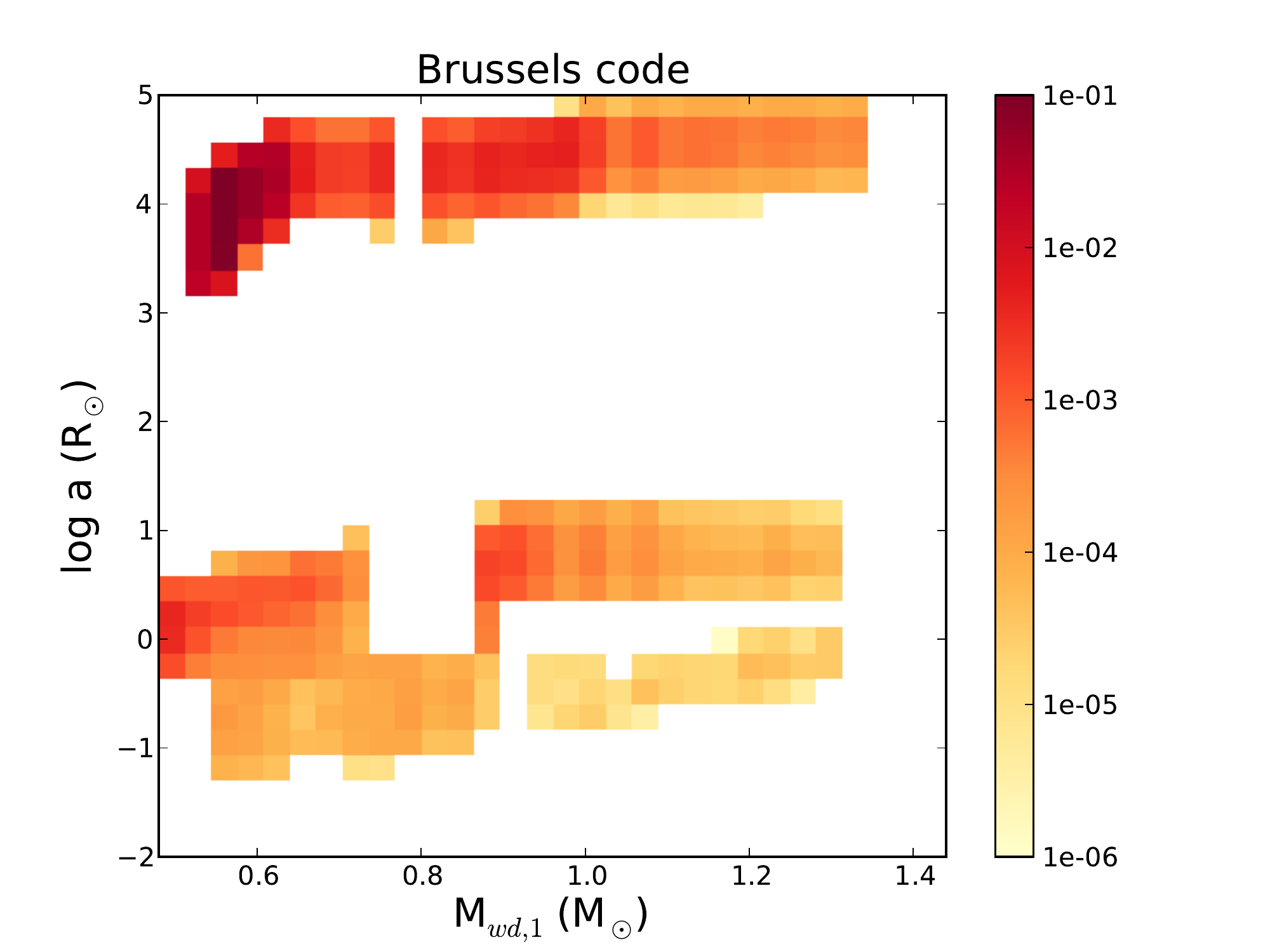}
\caption{Mass of the primary WD versus the separation at the moment of formation of the second WD, given by the different BPS codes. The numbers of the systems are normalized to the total number of the systems forming 2 WDs.}
\label{Sec}
\end{figure}

\begin{itemize}

\item\textbf{Similarities:}
The results of single WDs with a non-degenerate companion are very similar. At time of formation of the WD the same regions in separation and mass can be distinguished. In the population of double WDs three main regions can be distinguished and are similar in the results of the different groups (see Fig.~\ref{Sec}). A region with log $a \gtrsim$ 3, which are the non-interacting systems. A region with log $a \lesssim$ 1.5, which are systems resulting from a first mass transfer phase, which can be stable or unstable, and a second phase which is unstable. In the results of Binary$\_$c and SeBa there is a region around log $a$ equal to 1, which are systems resulting from a first phase of unstable mass transfer and a second phase which is stable. The percentage of double WD systems  compared to the simulated systems is similar for the different groups, namely 9\% for Binary$\_$c, for SeBa 10\% and 14\% for the Brussels code. 

\item\textbf{Differences:}
The predicted population of binary WDs are similar (see Fig.~\ref{Sec}), however small differences in mass and separation can be noticed. These are due to variations in the input physics: initial-final mass relation, the criterion to determine whether mass transfer continues stable or unstable, the exact duration and rate of mass transfer, the rate and angular momentum loss of the wind and the prescriptions to describe He-star evolution.
\end{itemize}

\paragraph{Conclusion}
Various BPS codes give similar results for the formation of binary WDs, when the assumptions are equalized. The differences in their results are not due to numerical differences, but can be explained by deviations in the input physics.

\bibliography{lit}

\end{document}